\documentclass[twocolumn]{revtex4-2}

\usepackage{float}

\usepackage{graphicx}
\usepackage{dcolumn}

\usepackage{bm}
 \usepackage{hyperref}
 \hypersetup{
     colorlinks=true,
     linkcolor=blue,
     filecolor=blue,
     citecolor = blue,      
     urlcolor=blue,
     }

\usepackage{tikz,xcolor,hyperref}

\definecolor{lime}{HTML}{A6CE39}
\DeclareRobustCommand{\orcidicon}{%
	\begin{tikzpicture}
	\draw[lime, fill=lime] (0,0) 
	circle [radius=0.16] 
	node[white] {{\fontfamily{qag}\selectfont \tiny ID}};
	\draw[white, fill=white] (-0.0625,0.095) 
	circle [radius=0.007];
	\end{tikzpicture}
	\hspace{-3.5mm}
}

\foreach \x in {A, ..., Z}{%
	\expandafter\xdef\csname orcid\x\endcsname{\noexpand\href{https://orcid.org/\csname orcidauthor\x\endcsname}{\noexpand\orcidicon}}
}

\usepackage{lineno}    

\begin{document}

\preprint{APS/123-QED}

\title{Ultra-Spatiotemporal Light Confinement in Dielectric Nanocavity Metasurfaces}

\author{Xia Zhang\orcidA{}}
\email{xzhang@tcd.ie}
\author{A. Louise Bradley\orcidB{}}%
 \email{bradlel@tcd.ie}
\affiliation{%
School of Physics, CRANN and AMBER, Trinity College Dublin, Dublin, Ireland
}%


\begin{abstract}

Spatiotemporal light confinement plays a crucial role in many branches of nanophotonics.
Here we propose a concept of quasi-bound states in the continuum gap cavity and reveal that ultra spatiotemporal confinement in free-space can be realized . By introducing an asymmetric air slot in a nanodisk resonator, an ultra-high quality factor $Q \sim 10^6$, accompanying an ultra-small effective mode volume, $V_m \sim 10^{-2}$ $(\lambda/n)^3$, is achieved resulting in a Purcell factor of $10^6$  $(\lambda/n)^{-3}$ in the visible wavelength range. The toroidal dipole and slot-effect drive the electric and magnetic field concentration in the air gap with a generated vortex polarizing electric field. Our study provides a new doorway for engineering light-matter interaction, such as Purcell factor enhancement, room temperature strong coupling, nonlinear nanophotonics and also may inspire studies on vortex beam generation and topological photonics.

\end{abstract}

\maketitle



Engineering single-photon emitter coupling with a resonant cavity is a fundamental topic in quantum electrodynamics \cite{vahala2003optical}. The coupling efficiency is predominantly determined by a cavity’s ability to spatially concentrate the field in a small volume and to suppress the dissipation of photons. In the weak coupling regime, these factors are quantified by the Purcell factor as $P_F\propto Q/V_{m}$ \cite{robinson2005ultrasmall}. In the strong coupling regime, these factors are expressed in terms of the coupling strength $g$, where $g = \sqrt{N} \bm{\mu}_e \cdot \bm{E} \propto \bm{\mu}_e \sqrt{N/V_m}$, with $\bm{\mu}_e$ the emitter's dipole moment and $\bm{E}$ the electric field confined in the mode volume, $N$ is the number of the emitters placed inside or in the near field of the cavity \cite{torma2014strong, hugall2018plasmonic,lawless2020influence}. $Q$ is the quality factor of the mode and $V_m$ is the effective mode volume. Optimizing $Q/V_{m}$ has been a long-time pursuit to realize efficient light-matter interaction, and has been an intensive topic of research over the last three decades due to not only fascinating fundamental properties from a quantum optics point of view, such as high-temperature Bose–Einstein condensation \cite{kasprzak2006bose}, but also for technological applications such as low threshold polariton  lasing \cite{mckeever2003experimental}, single photon switches \cite{volz2012ultrafast}, nonlinear nanophotonics \cite{koshelev2020subwavelength} and quantum information \cite{van2013photon}.

To date high $Q/V_m$ ratios have been achieved in a relatively large-scale dielectric Fabry-Perot microcavity \cite{andreani1999strong,askitopoulos2011bragg, min2009high,liu2015strong,hu2018experimental}, miniature plasmonic nanocavity \cite{zengin2015realizing,schuller2010plasmonics, chikkaraddy2016single} and a hybrid system \cite{todorov2009strong}. Generally, the plasmonic cavity outperforms a dielectric cavity to enable  efficient light-matter interaction due to the ultra-confined field strength and sub-diffraction-limited effective mode volumes $V_{m}\sim 10^{-4}  (\lambda/n)^3$ \cite{hugall2018plasmonic}. However, energy dissipation is an inevitable issue in plasmonic nanoresonator, and consequently with lower achievable $Q$ factors, typically in the range $\sim$10-100 \cite{chikkaraddy2016single,carlson2020dissipative}. Oppositely, a dielectric cavity can provide a large $Q$ factor ($\sim10^5-10^6$) employing Bragg mirror or whispering gallery mode configurations but cannot efficiently spatially trap the electric field in a small mode volume in air. Conventional dielectric resonators have mode volumes $\sim(\lambda/2)^3$ \cite{schuller2010plasmonics}. Additionally, most of the field concentration exists inside the dielectric resonator rather than in the near field \cite{evlyukhin2016optical, he2018toroidal, hugall2018plasmonic}. It is therefore challenging to place an individual dipole emitter in the free-space dielectric hot-spot for efficient coupling. Hybrid plasmonic-dielectric structures have also been considered in order to compensate their strength and weakness and to further optimize $Q/V_{m}$ \cite{luo2015chip}, though requiring complex numerical optimization and technical fabrication. An intriguing question arises: is it possible to create a free-space dielectric hot-spot with ultra-spatiotemporal field confinement? The dielectric hot-spots may inspire rich branches of physics and provide a new route for potential applications including  nonlinear optics \cite{koshelev2020subwavelength},  sensing \cite{maksimov2020refractive},  hyper-spectral  imaging \cite{tittl2018imaging},  emission control \cite{yuan2017strong}, topological photonics \cite{doeleman2018experimental,wang2020generating} and quantum technologies \cite{rose2017coherent}.

Regarding temporal confinement, recent developments with dielectric nanoresonators have revolutionized optical cavity design by employing BIC \cite{koshelev2018asymmetric, sadrieva2019multipolar, liu2019high}. BIC, first discovered in quantum systems by von Neumann and Wigner \cite{friedrich1985interfering} and later proved from Maxwell’s theory \cite{marinica2008bound,ndangali2010electromagnetic}, is in principle wave solutions that are embedded in a radiative continuum and ideally non-radiative, thus with an infinitely high $Q$ but inaccessible. However, the eigenmode becomes accessible when the symmetry is broken via an oblique angle of incidence or geometric symmetry breaking \cite{koshelev2018asymmetric, liu2019high,doeleman2018experimental}, where the BIC becomes a quasi-BIC and has an accessible but still finite-high $Q$ \cite{hsu2016bound,liu2019high}. Regarding spatial confinement, the toroidal dipole has shown the exceptional capability to realize field confinement \cite{he2018toroidal}. Further inspired by conventional photonic crystal (PhC) nanobeam cavity design for field enhancement employing slot effect and anti-slot effect \cite{robinson2005ultrasmall,ryckman2012low, hu2016design} or refractive index contrast, we propose the concept of a quasi-BIC gap cavity, where an asymmetric air slot is introduced in a dielectric resonator, which exhibits a toroidal dipole resonance to realize spatiotemporal light confinement and as a testbed to explore capabilities of manipulating light-matter coupling.


\begin{figure}
\includegraphics[width=1\linewidth]{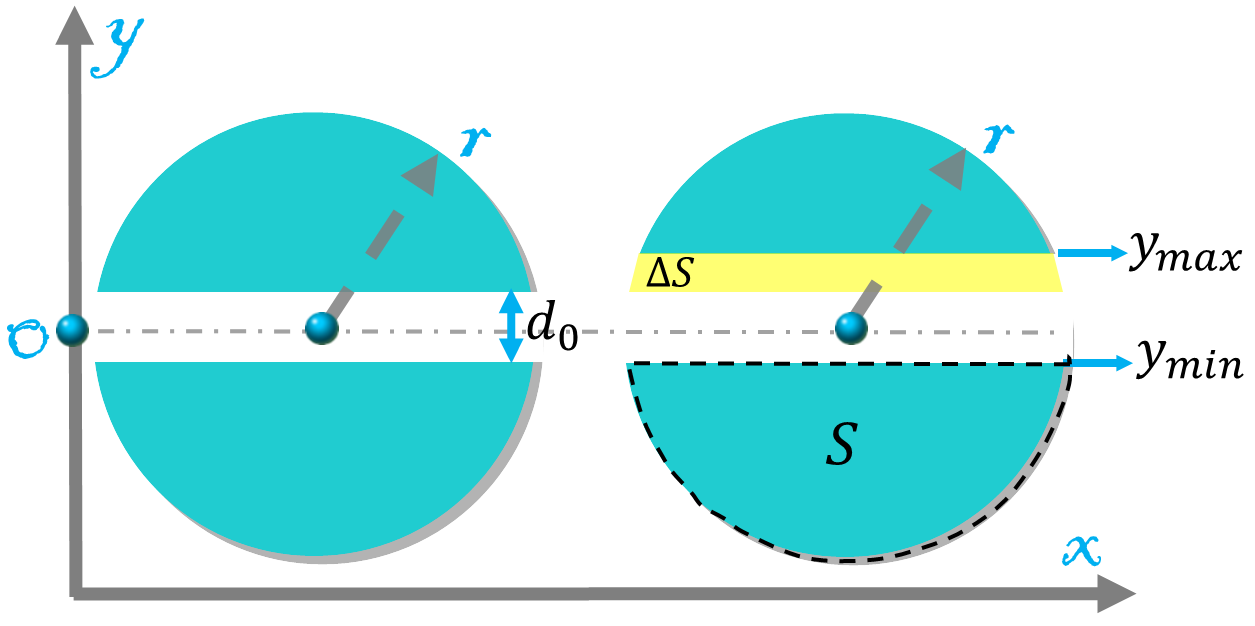}
\caption{\label{fig:sch} Schematic graph of a unit cell : a dielectric nanodisk with an inserted slotted air gap. The coordinate origin is shown as $O$. The position of the air gap is denoted as $y_{max}$ and $y_{min}$. $d_0$ refers to the gap width of the symmetric nanodisk. $r$ refers to the nanodisk radius. By varying $y_{max}$ only with a fixed $y_{min}$, symmetry is broken along $y$ direction. $S$ denotes the half disk area minus the half air gap area without symmetry breaking, illustrated as the black dash line. $\Delta S$ denotes the symmetry breaking by only varying $y_{max}$ for reduced area of upper part and the corresponding asymmetry parameter is $\alpha=\Delta S/S$.}
\end{figure}

\begin{figure}
\includegraphics[width=1\linewidth]{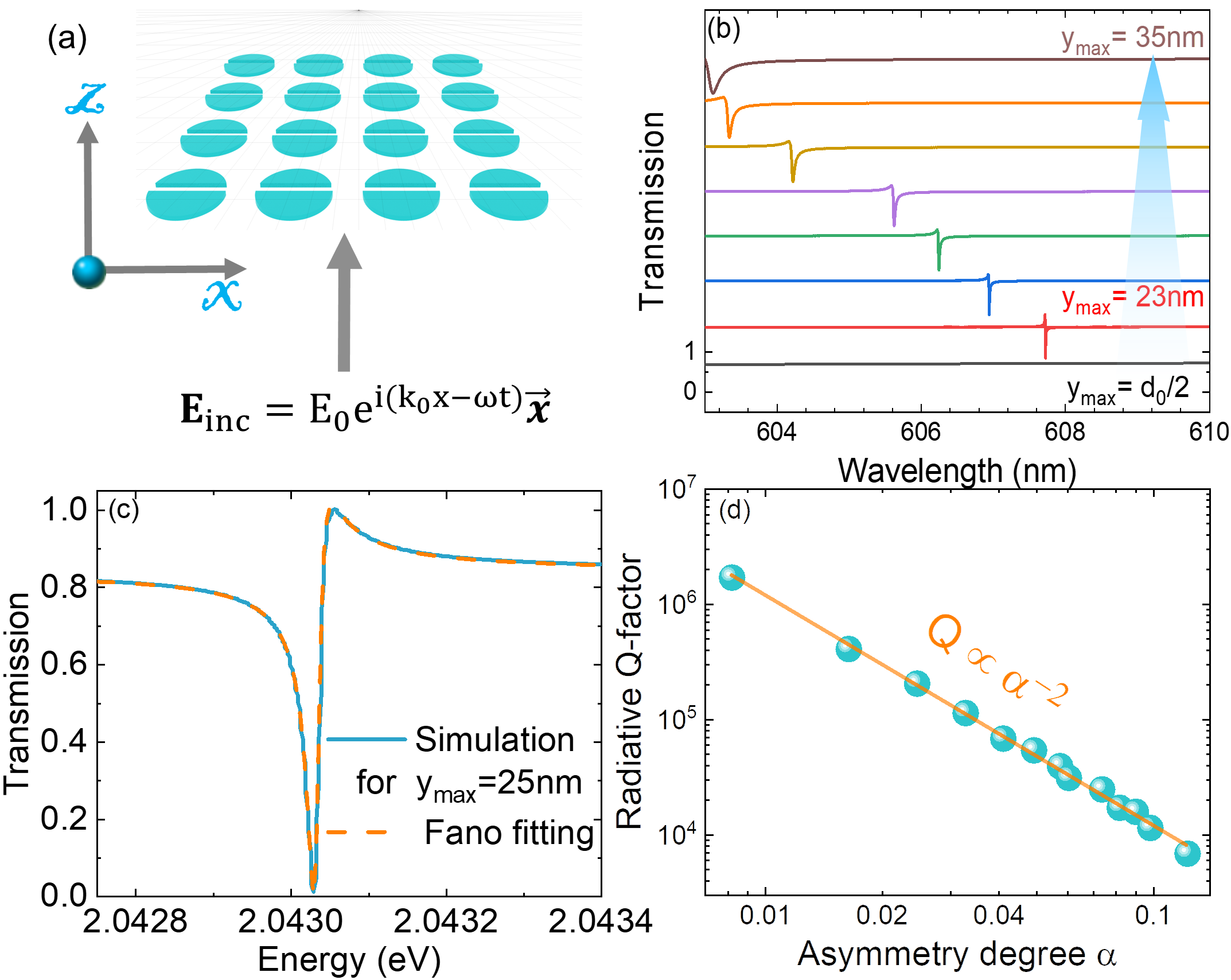}
\caption{\label{fig:nature}(a) Schematic of the designed metasurface in free-space, which is normally illuminated by a wave, $\bm{E}_{inc}=E_0e^{ik_0 z-\omega t}\bm{x}$, polarizing along the slotted air gap. (b) Simulated transmission spectra of BIC cavity ( $|y_{max}|=|y_{min}|=d_0/2$ = 20 nm) and quasi-BIC cavities (fixed $y_{min}$ = -20 nm and varying $y_{max}$: shown for $y_{max}$ ranging from 23 nm to 35 nm with a step of 2 nm, $r$ = 180 nm, $h$ = 100 nm, $p_x = p_y = 500$ nm). (c) A Fano fit of the transmission spectra for $y_{max}$ = 25 nm. (d) The calculated radiative $Q$ factor vs asymmetry parameter $\alpha = \Delta S/S$ (log-log scale). The shown minimum asymmetry factor is for $y_{max}$ = 21 nm with $\alpha$ = 0.008 and a gap width $d$ = 41 nm. }
\end{figure}

The designed cavity is illustrated in Fig.~\ref{fig:sch}. An air slotted gap is introduced within the disk and a gap cavity is created. $y_{max}$ and $y_{min}$ represent the positions of the gap and the width $d = y_{max}-y_{min}$. The designed nanocavity is symmetric while $|y_{max}|=|y_{min}|$ where $d_0$ denotes the corresponding width. It becomes asymmetric in $y$ direction when $|y_{max}|\neq|y_{min}|$. To achieve symmetry breaking, $y_{min}$ is fixed and $y_{max}$ is varied. $\Delta S$ refers the reduced disk area by increasing the cavity width and can be approximated as a trapezoidal area with $\Delta S= (\sqrt{r^2-y_{max}^2}+r)(y_{max}-d_0/2)$. The asymmetry parameter is defined as $\alpha = \Delta S/S$. $S$ denotes the half disk area minus half air gap area without symmetry breaking, which is illustrated as the black dash line in Fig.~\ref{fig:sch}.

Arranging the designed cavity in a periodic array constitutes a metasurface. The metasurface period is fixed as $p_x = p_y$ = 500 nm, the height of disk is $h$ = 100 nm and radius $r$ = 180 nm. The dielectric cavity metasurface explored here is made of titanium dioxide, whose permittivity is taken from the experimental data \cite{sarkar2019hybridized} and is lossless in the visible wavelength. Three-dimensional (3D) finite-difference time-domain (FDTD) simulations were employed to calculate the optical properties. The metasurface is illuminated by a normally incident wave, $\bm{E}_{inc}=E_0e^{ik_0 z-\omega t}\bm{x}$, which is polarised along the slotted air gap. The transmission spectra for the $y$-symmetric metasurface ($y_{min}$ = -20 nm, $y_{max}$ = 20 nm) and $y$-asymmetric metasurface ($y_{min}$ = -20 nm, varied $y_{max}$) can be seen in Fig.~\ref{fig:nature} (b). Due to symmetry breaking along $y$ direction, a sharp eigenmode appears in each transmission spectra with varying linewidth, each of which can be fitted by Fano formula, an example is shown in Fig.~\ref{fig:nature} (c), see Supplementary Material \cite{2021PRsupp} for the fit details and $Q$ calculation. The calculated $Q$ factor vs asymmetry parameter $\alpha$ can be seen in Fig.~\ref{fig:nature} (d), where an inverse quadratic law ($Q\propto \alpha^{-2}$) is met. This is consistent with the results reported by Koshelev et al. \cite{koshelev2018asymmetric} and thus demonstrates the designed gap cavity is quasi-BIC in nature. More physics of $Q$ evolution by multipolar contributions can be seen in Supplementary Material \cite{2021PRsupp}. In this work, the $y$-symmetric cavity is referred to as the BIC cavity, where the infinitely-high $Q$ is inaccessible and $y$-asymmetric cavity is referred to as the quasi-BIC cavity, which has the accessible and high $Q$. The eigenmode wavelength  of the quasi-BIC cavity denotes the wavelength close to the transmission minimum and the maximum electric field can be achieved.

\begin{figure*}[t]
\includegraphics[width=1\linewidth]{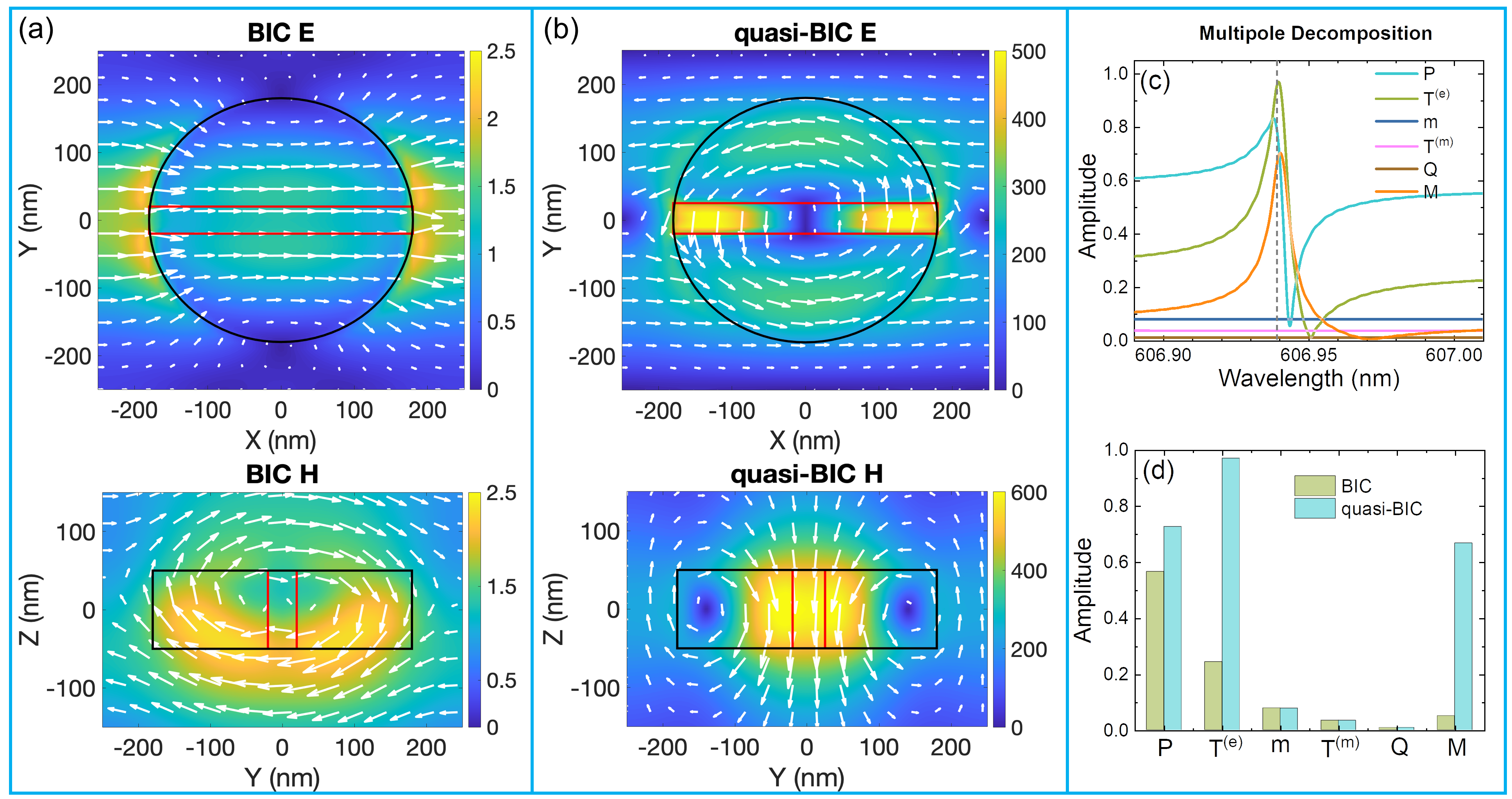}
\caption{\label{fig:field}The calculated amplitude ratio of electric field and magnetic field maps at the eigenmode wavelength (both inspected at 606.94 nm), corresponding to $|\mathbf{E}/\mathbf{E}_0|$ in $x-y$ plane at $z$ = 0 and $|\mathbf{H}/\mathbf{H}_0|$ in $y-z$ plane at $x$ = 0 for (a) BIC cavity ($|y_{min}|=|y_{max}|$ = 20 nm) and (b) quasi-BIC cavity ( $y_{min}$ = -20 nm, $y_{max}$ = 25 nm). The black line denotes the cavity and the red line denotes the slotted air gap. (c) The calculated scattering amplitude of multipole modes contributing to the far-field scattering, including the electric dipole, $\mathbf{P}$, electric toroidal dipole $\mathbf{T^{e}}$, magnetic dipole $\mathbf{m}$, magnetic toroidal dipole $\mathbf{T^{m}}$, electric quadrupole, $\mathbf{Q}$ and magnetic quadrupole $\mathbf{M}$. The black dash line illustrates the inspected eigenmode wavelength at 606.94 nm, see wavelength-dependent field amplitude in Supplemental Material \cite{2021PRsupp}. (d) The comparison of scattering amplitudes of the BIC cavity and quasi-BIC cavity at 606.94 nm.}
\end{figure*}


\begin{figure}[ht]
\includegraphics[width=0.9\linewidth]{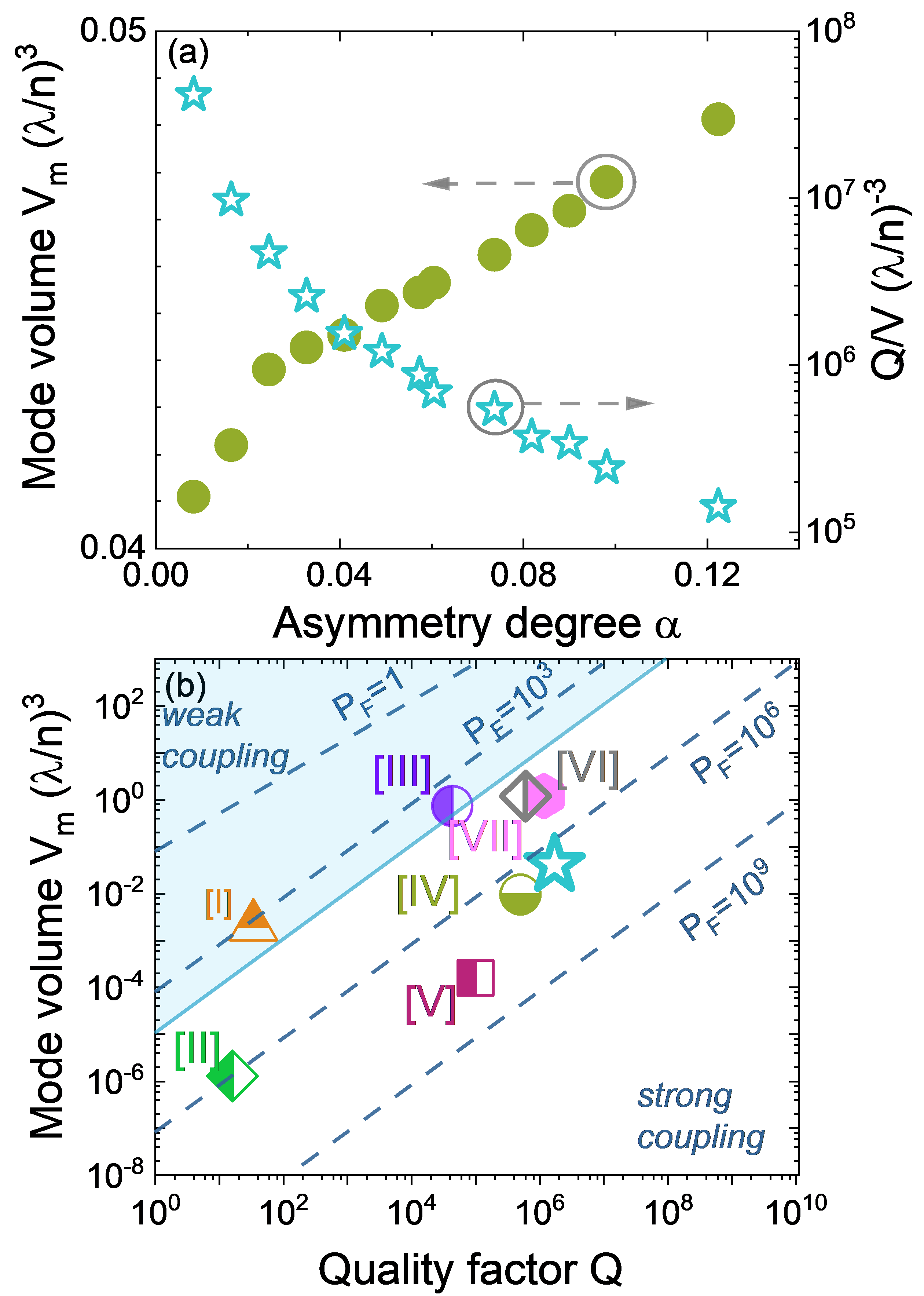}
\caption{\label{fig:vm}(a) The calculated effective mode volume, $V_m$, and $Q/V_m$ at the eigenmode wavelength versus asymmetry parameter $\alpha$. (b) Comparison with state-of-the-art plasmonic or PhC optical cavities with various geometries under ambient conditions: including (I) plasmonic dimer \cite{carlson2020dissipative}, (II) plasmonic nanoparticle-on-mirror \cite{chikkaraddy2016single}, (III) 2D PhC with L3 effect \cite{akahane2003high}, (IV) dielectric slotted nanobeam \cite{quan2010photonic}, (V) dielectric bowtie nanobeam \cite{hu2018experimental}, (VI) slotted nanobeam \cite{seidler2013slotted} and (VII) 2D PhC heterojunction \cite{song2005ultra}. 
The star denotes this work for a quasi-BIC gap dielectric nanocavity, with the position corresponding to the theoretical calculated $Q$ factor and effective mode volume. The dash lines correspond to Purcell factors, $P_F$, of different values. The shaded area represents the room temperature single molecule weak coupling regime and unshaded area represents the regime for strong coupling. }
\end{figure}

Since the proposed quasi-BIC gap cavity has shown strong temporal confinement, we proceed to explore the spatial field confinement. The amplitude ratios of the electric field, $|\mathbf{E/E_0}|$ in the $x-y$ plane at $z$ = 0 and magnetic field, $|\mathbf{H/H_0}|$ in the $y-z$ plane at $x$ = 0 are shown in Fig.~\ref{fig:field}. The corresponding field vectors are also shown. As can be seen from Fig.~\ref{fig:field} (a), there are only 
$\sim$ 2.4 fold electric and magnetic field enhancement inside and in the near field of the BIC cavity (gap width $\sim$ 40 nm). There is a linear current displacement along $x$, resulting in an azimuthal magnetic field as seen in BIC $\mathbf{H}$ map. The generated magnetic field is mostly confined in a torus-like shape, which threads the two symmetric parts together by the inward (along $x$) displacement current. Regarding the relatively low $\textbf{E}$ and $\textbf{H}$ field enhancement seen here, it is worth mentioning that a plasmonic gap cavity shows a large electric field when the incident light is polarized across the gap rather than along the gap employing the 
interaction between adjacent parts of a plasmonic dimer resonator separated by a small gap or field enhancement arising due to the sharp features of the resonator, the so-called lightning rod effect \cite{schuller2010plasmonics}. The localized field enhancement for light polarized along and across the gap in an symmetric dielectric dimer can be found elsewhere \cite{bakker2015magnetic}, where a maximum of only $\sim$ 10 fold is achieved at visible wavelengths. However, as can be seen from Fig.~\ref{fig:field} (b),
a maximum of $\sim$ 500 fold $|\mathbf{E/E_0}|$ and $\sim$ 600 fold $|\mathbf{H/H_0}|$ enhancement is generated within the air gap (gap width $\sim$ 45 nm) due to slot-effect, more details can be seen in the Supplemental Material \cite{2021PRsupp}.

Furthermore, an distinct in-plane vortex electric field ($x-y$) is formed in quasi-BIC cavity, with a corresponding azimuthally polarized or vortex displacement current. A similar BIC-related vortex beam was recently reported employing symmetry breaking by oblique angle of incidence \cite{wang2020generating}. Here we find the vortex beam can be realized under normal angle of incidence but with slightly geometric symmetry breaking. Moreover, clockwise and anti-clockwise magnetic field loops in $y-z$ plane at the quasi-BIC resonance are formed in the magnetic field pattern, indicating the existence of the toroidal dipole. Toroidal dipole BIC has exceptional capability to confine the field inside the resonator \cite{he2018toroidal}. In order to probe further the physics behind the large field confinement, multipole decomposition is employed \cite{evlyukhin2016optical,li2018origin}. The 3D electric field distribution used for calculation is within one unit cell length in $x-y$ direction and within the height of of cavity in $z$ direction \cite{basharin2015dielectric}. The amplitudes for all multipolar modes are shown in Fig.~\ref{fig:field} (c), including electric dipole $\mathbf{P}$, electric toroidal dipole $\mathbf{T^{(e)}}$, magnetic dipole $\mathbf{m}$, magnetic toroidal dipole $\mathbf{T^{(m)}}$, electric quadrupole, $\mathbf{Q}$ and magnetic quadrupole $\mathbf{M}$. For a direct comparison, the amplitude of each for BIC cavity and quasi-BIC cavity are all presented in Fig.~\ref{fig:field} (d). It is clear that electric dipole $\mathbf{P}$ dominates the far-field scattering of the BIC cavity. However, electric toroidal dipole $\mathbf{T^{(e)}}$ dominates and more pronounced magnetic quadrupole appears in the quasi-BIC cavity, which corresponds with the magnetic field loop observed in the $\mathbf{H}$ field map shown in Fig.~\ref{fig:field} (b).  See Supplementary Material
\cite{2021PRsupp} for more information on field map and multipolar contributions vs the asymmetry factor.

The effective mode volume is further analyzed to probe the potential capabilities for positioning a single molecule in the air gap to enable strong coupling. The dimensionless effective mode volume for a dielectric cavity is given by the ratio of the total electric energy to the maximum electric energy density \cite{ kristensen2012generalized}
\begin{equation}
V_m=\frac{\int_V{\epsilon(\bm{r})|\bm{E(r)}|^2}dV}{\epsilon (\vec{\bm{r}}_{max})max [|\vec{\bm{E}}(\vec{\bm{r}}_{max})|^2]}[\frac{n(\vec{\bm{r}}_{max})}{\lambda}]^3
\end{equation}
where $\vec{\bm{r}}_{max}$ denotes the position of the peak magnitude of the electric field $|\bm{E}|$ and $n (\vec{\bm{r}}_{max})$ is the corresponding index of refraction.  $\lambda$ denotes the free-space wavelength. The electric field is integrated over one unit cell by corresponding $V =p_x\times p_y\times h$. The same approach calculating $V_m$ for the metasurface can be found elsewhere \cite{liang2020bound, gupta2020terahertz}.

As can be seen in Fig.~\ref{fig:vm} (a), an order of $V_m\sim10^{-2}(\lambda/n)^3$ is achieved for the quasi-BIC gap cavity. For a direct comparison, the calculated $Q$, $V_m$ and Purcell factor, $P_F=3Q/(4\pi^2V_m)$ are shown in Fig.~\ref{fig:vm}  (b) for state-of-the-art plasmonic and PhC optical cavities with various geometry, including [I] plasmonic dimer \cite{carlson2020dissipative}, [II] plasmonic nanoparticle-on-mirror \cite{chikkaraddy2016single}, (III) 2D PhC with L3 effect \cite{akahane2003high}, (IV) dielectric slotted nanobeam \cite{quan2010photonic}, (V) dielectric bowtie nanobeam \cite{hu2018experimental}, (VI) slotted nanobeam \cite{seidler2013slotted} and (VII) 2D PhC heterojuction \cite{song2005ultra}. 
This work with $y_{max}=21$ nm, corresponding to an asymmetry factor $\alpha = 0.008$, is denoted by the star, where $P_F$ can reach a value of the order of $10^6$, which outperforms most of the reported $P_F$ of state-of-the-art optical cavities working under normal incidence to the best of our knowledge. The theoretical $Q$ can be infinitely high when the asymmetry factor $\alpha$ is infinitely small, however technical fabrication with sub-nanometer precision is challenging. Although the bowtie nanobeam cavity at a $\sim$12 nm tip (partially in air and partially in silicon) has a reported higher $P_F$ than this work \cite{hu2018experimental}, the proposed quasi-BIC gap cavity here has a width of 41 nm with both electric and magnetic field concentration in air gap in each unit cell of metasurface, which enables greater accessibility, easier technical fabrication and flexibility in positioning individual or ensemble dipole emitters. For room temperature single molecule with the typical scattering rate of dipole emitter $\sim k_B T$, strong coupling requires $V_m$ less than $10^{-5}$ \cite{chikkaraddy2016single}, which is represented as the line in Fig.~\ref{fig:vm} (b) separating two regions, where below the line (clear area) represents the strong coupling regime and above the line (shaded area) is the weak coupling regime. It is clear that the proposed quasi-BIC gap cavity has an even higher Purcell factor than the sub-nanometer plasmonic nanoparticle-on-mirror system and thus provides an alternative platform to enable room-temperature strong coupling.


To conclude, we address the intriguing question that whether a free-space dielectric hot-spot exists, analogous to plasmonic near-field hot-spot.  The concept of quasi-BIC gap nanocavity is proposed with an inserted asymmetric slotted air gap with a generating symmetry-breaking perturbation. The quasi-BIC nature is proven, with a strong temporal confinement. A $Q$ factor up to $10^{6}$ for a gap $d =$ 41 nm is achieved, with corresponding asymmetry degree $\alpha = $ 0.008. A dielectric hot-spot is generated within the asymmetric slotted air gap, with simultaneous squeezing of electric magnetic field at the eigenmode wavelength. The dominant toroidal dipole resonance and slot-effect drive the field concentration.
Furthermore, the effective mode volume is of the order of $10^{-2}(\lambda/n)^3$ at visible wavelengths, two order of magnitude smaller than traditional nanobeam cavities. A Purcell factor of the order of $10^{6}$ is theoretically realized in a simple dielectric metasurface geometry under normal light of incidence, which can be employed for room-temperature single molecule strong coupling. Plasmonic cavities with near-field hot-spots have served as a platform for manipulating light-matter interaction over the last few decades, the proposed quasi-BIC gap cavity with dielectric hot-spots provides an exciting alternative platform for cavity QED, lasing, optical trapping, quantum light manipulation and is expected to inspire richer branches of physics, such as topological photonics and vortex beam generation in metaoptics.

\begin{acknowledgments}
We wish to acknowledge the support of Science Foundation Ireland (SFI) under Grant Number 16/IA/4550.
\end{acknowledgments}



\nocite{*}

\bibliography{apssamp}

\textbf{Supplementary Material}

\section{Fano fit of the transmission spectra and Q calculation}

The simulated transmission spectra displays a Fano profile and can be phenomenologically fitted by $T(E)=T_0+A_0\frac{[q+2(E-E_0)/\Gamma]^2}{1+[2(E-E_0)/\Gamma]^2}$ \cite{fang1998determination},
where $E_0$ is the resonance energy and $E = hc/\lambda$, $\Gamma$ is the resonance energy linewidth, $\lambda$ is the free space wavelength, $T_0$ is the transmission offset, $A_0$ is the continuum-discrete coupling constant, and $q$ is the Breit-Wigner-Fano parameter determining the asymmetry of the resonance profile. The $Q$ factor of the Fano resonance is evaluated by $Q=E_0/\Gamma$.

\section{Quality Factor Evolution with Multipolar Analysis}

According to Ref. \cite{koshelev2018asymmetric,evlyukhin2016optical}, the quality factor of a quasi-BIC gap cavity, $Q$, is related to multipolar contributions and can be approximated as
\begin{widetext}
\begin{equation}
\label{qf}
Q=E_0/\Gamma=\omega_0/\gamma=\frac{2A}{k_0[|p_x-\frac{1}{c}m_{y}+\frac{ik_0}{6}Q_{xz}|^2+|p_y+\frac{1}{c}m_{x}+\frac{ik_0}{6}Q_{yz}|^2]}
\end{equation}
\end{widetext}
where $p_{x,y}=P_{x,y}+ikT_{x,y}$, is the total electric dipole, $\textbf{P}$, $\textbf{T}$, $\textbf{m}$ and $\textbf{Q}$ are the electric dipole, toroidal dipole, magnetic dipole and electric quadrupole moments in the irreducible representations defined as in Ref. \cite{evlyukhin2016optical}. $A$ is the area of one unit cell. $k_0$ is the free-space wavevector. $E_0$, $\omega_0$ are the resonance energy and frequency, $\Gamma$, $\gamma$ are the corresponding energy width and frequency width, respectively. Under $x$-polarized light normally incident light, Eq.~\ref{qf} can be simplified as $Q = 2A/[k_0p_x^2] = 2A/[k_0\alpha^2 p_0^2] \propto \alpha^{-2}$, where $p_x=\pm\alpha p_0$, $\alpha$ is the geometric asymmetry parameter of the quasi-BIC metasurface, $p_0$ is the total electric dipole moment in the BIC metasurace without symmetry-breaking. In the asymptotic region for $\alpha\rightarrow0$, the quality factor of the transmission coefficient $Q \rightarrow \infty$, where the quasi-BIC cavity evolves to an ideal BIC cavity.

\section{Field Concentration within the Air Gap:
Slot Effect}

Multipolar mode contributions as well as the slot effect explain the field concentration within the air gap. 
BIC and quasi-BIC metasurfaces correspond to different symmetry groups, resulting in the excitation of distinct dominant multipolar modes under the same illumination with $x$ polarized waves at normal incidence, which determines the orientation of the electric field and magnetic field vectors. An abrupt refractive index discontinuity along the field vector results in the field concentration within the low-refractive index region \cite{robinson2005ultrasmall, choi2017self, hu2018experimental}.
As an example in Fig.~\ref{fig:point} at the illustrated position with the maximum field amplitude, the orientation of electric field vector is perpendicular to the interface and the orientation of magnetic field vector is parallel along the interface. According to Maxwell's electromagnetic boundary conditions at the dielectric and air interface:  

\begin{equation}
\label{Eperp}
    \epsilon_l E_{l\perp} =D_{l\perp}=D_{h\perp}=\epsilon_h E_{h\perp}
\end{equation}

\begin{equation}
\label{Epare}
     E_{l||} = E_{h||}
\end{equation}

\begin{equation}
\label{Hperp}
    \mu_l H_{l\perp} =B_{l\perp}=B_{h\perp}=\mu_h H_{h\perp}
\end{equation}

\begin{equation}
\label{Hpare}
     H_{l||} = H_{h||}
\end{equation}
where $\epsilon_l$, $\epsilon_h$ represents the low and high permittivity forming the interface. Corresponding, $\mu_l$, $\mu_h$ represent the low and high permeability forming the interface. A subscript $||$ and $\perp$ refers to parallel and normal components. Due to the dominant normal electric field and parallel magnetic field components at the interface at the position shown in Fig.~\ref{fig:point}, the Eq.~\ref{Eperp} and Eq.~\ref{Hpare} explain the field concentration. As can be seen close to the air-dielectric interface, the electric field within the air gap (low refractive index, $\epsilon_l$ = 1) is roughly two-fold that in the dielectric area (high refractive index, $\epsilon_h\approx$ 4.5). The magnetic field amplitude is roughly the same at the interface, verified by Eq.~\ref{Hperp}. The electric field amplitude ratio is not only determined by the ratio of the permittivity, but also strongly dependent on the geometry of the air gap, such as bowtie tip air gap shows a much higher field concentration than that of the circular and slotted air gap \cite{hu2018experimental}. Furthermore, the field amplitude concentration of the designed quasi-BIC cavities also depends on the multipolar modes and slot width, as can be seen from Fig.~\ref{fig:field}.

\newpage
\renewcommand\thefigure{S\arabic{figure}}
\begin{figure*}[h]
\setcounter{figure}{0}
\includegraphics[width=1\linewidth]{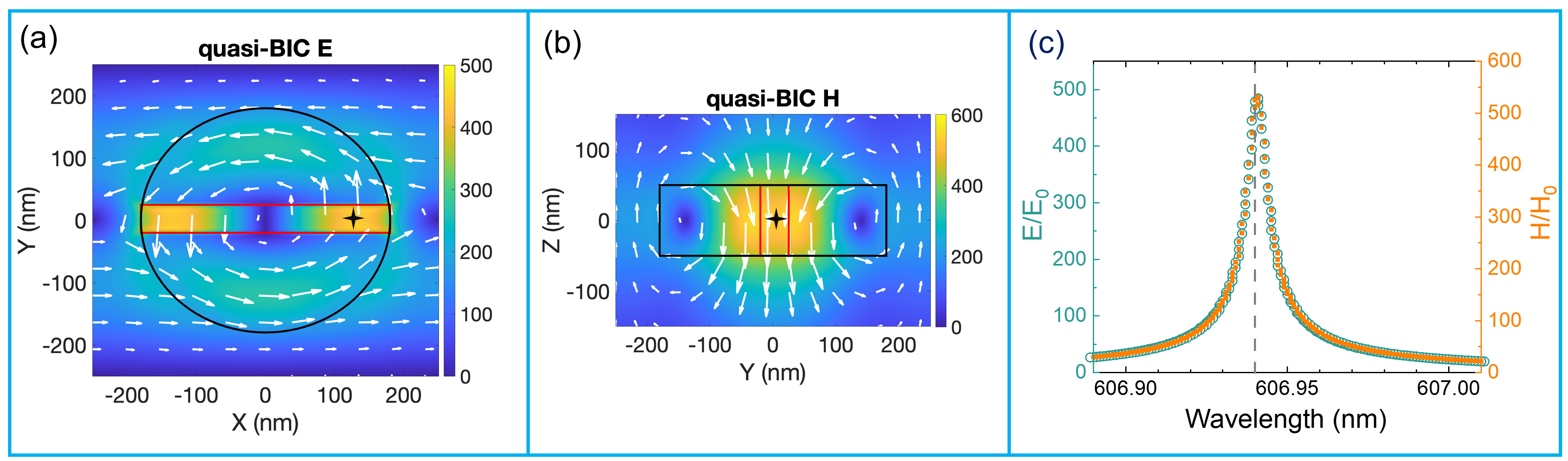}
\caption{\label{fig:point} The calculated amplitude ratio of electric field (a) and magnetic field (b) maps of quasi-BIC metasurface for $y_{min} = -20$ nm and $y_{max}= 25$ nm, where the black cross symbol $+$ represents where the point field monitor is placed ($x_0, y_0, z_0$ = [125 nm, 2.5 nm, 0]) with the achieved maximum electric field amplitude ratio. (c) The wavelength dependent E and H amplitude ratio at the position indicated in (a) and (b). }
\end{figure*}

\renewcommand\thefigure{S\arabic{figure}}
\begin{figure*}[h]
\includegraphics[width=1\linewidth]{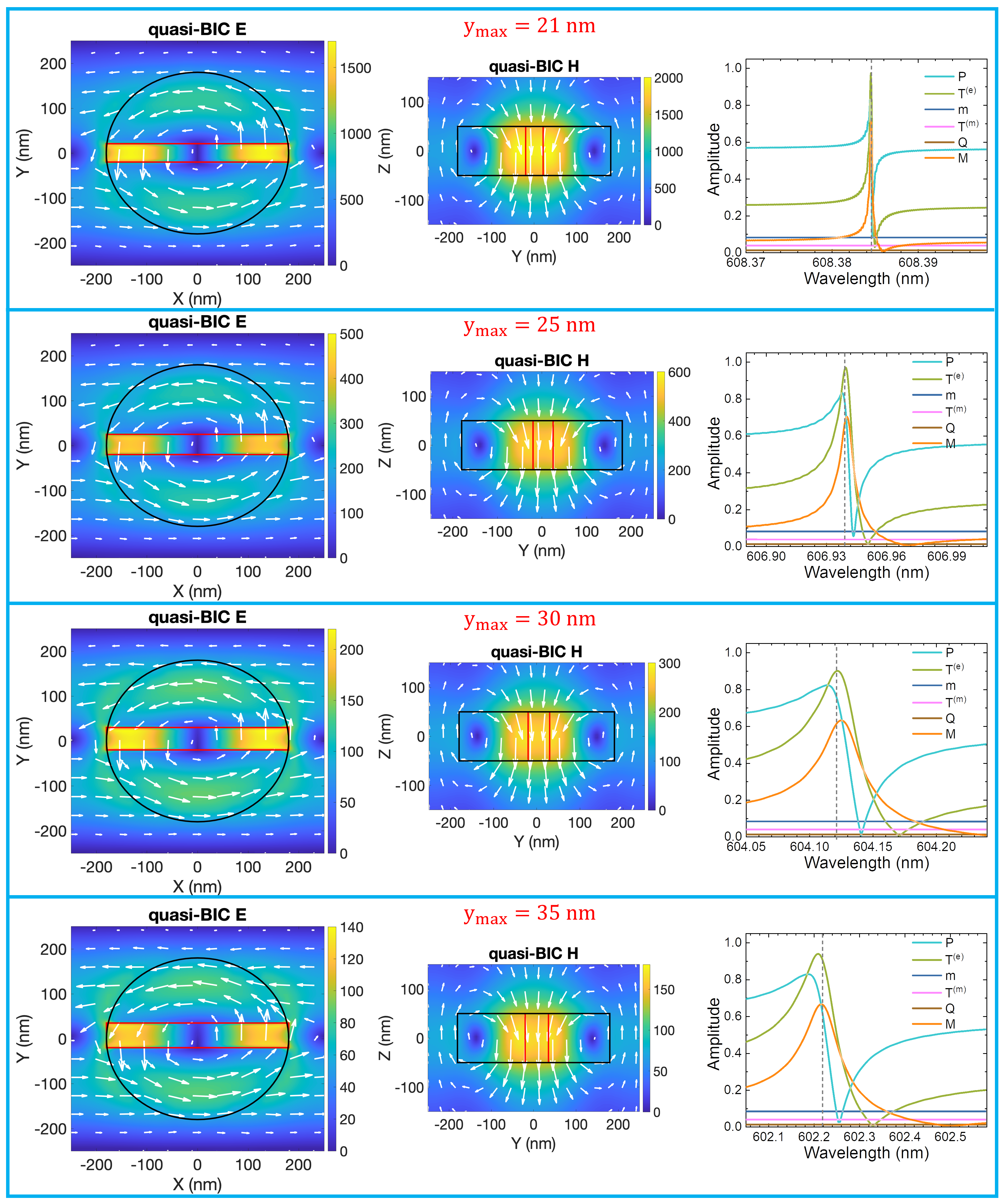}
\caption{\label{fig:field} The calculated amplitude ratio of electric field and magnetic field maps of quasi-BIC metasurface with varying asymmetry parameter.  The wavelength is selected at which the maximum electric field amplitude is achieved and are shown as the gray dash lines in the right. The results are shown for metasurfaces with a constant $y_{min} = -20$ nm and varied $y_{max}$ for 21 nm, 25 nm, 30 nm and 35 nm, corresponding to asymmetry parameter $\alpha$ of 0.0082, 0.041, 0.082, 0.1224 respectively. }
\end{figure*}

\end{document}